\documentclass[12pt]{iopart}

\usepackage{t1enc}
\usepackage{cite}

\usepackage{iopams}
\usepackage{amsfonts}
\usepackage{epsfig}

\newcommand{\w}{\boldsymbol{\omega}}

\newcommand{\met}{\boldsymbol{g}}
\newcommand{\bfe}{\boldsymbol{e}}
\newcommand{\der}{\,\mathrm{d}}

\newcommand{\deq}{:=}
\newcommand{\al}{\alpha}
\newcommand{\be}{\beta}
\newcommand{\bfT}{\boldsymbol{T}}
\newcommand{\bfu}{\boldsymbol{u}}
\newcommand{\sgn}{\mathrm{sgn}}
\newcommand{\ie}{i.e.}
\newcommand{\eg}{e.g.}
\newcommand{\tfrac}{\textstyle\frac}

\makeatletter
\newcommand{\artsectnumbering}{%
\@addtoreset{equation}{section}
\renewcommand{\theequation}{\thesection.\arabic{equation}}}
\makeatother

\begin{document}
\jl{6}
\title[Construction of LRS perfect fluids]%
{Invariant construction of solutions to Einstein's
    field equations -- LRS perfect fluids II}

\author{M Marklund\footnote[1]{E-mail address: %
        mattias.marklund@physics.umu.se} and M
      Bradley\footnote[2]{E-mail address:
      michael.bradley@physics.umu.se}}  
\address{Department of Plasma Physics, Ume{\aa} University, S-901 87
    Ume{\aa}, Sweden}

\artsectnumbering
 
\begin{abstract}
  The properties of LRS class II perfect fluid space-times
  are analyzed using the description of geometries in terms of 
  the Riemann tensor and a finite number of its covariant
  derivatives. In this manner it is straightforward to obtain the
  plane and hyperbolic analogues to the spherical symmetric case.
  For spherically symmetric static models the set of equations is
  reduced to the Tolman--Oppenheimer--Volkoff equation \emph{only}.
  Some new non-stationary and inhomogeneous solutions with shear,
  expansion, and acceleration of the fluid are presented.
  Among these are a class of temporally self-similar solutions with
  equation of state given by $p=(\gamma-1)\mu$, $1<\gamma<2$, and a
  class of solutions characterized by $\sigma=-{\tfrac 16}\Theta$.
  We give an example of a geometry where the Riemann tensor and the 
  Ricci rotation coefficients are not sufficient to give a complete 
  description of the geometry.
  Using an extension of the method, we find the full metric in
  terms of curvature quantities. 
\end{abstract} 

\submitted


\section{Introduction}

It is well known that manifolds with a metric structure, i.e., 
geometries, are locally completely determined by
a set consisting of the Riemann tensor and its covariant derivatives 
with respect to a frame with constant metric components $\eta_{ij}$,
\cite{Cartan, Brans, Karl}. This set will henceforth be called
\[
  R^{p+1} \deq \{R_{ijkl},R_{ijkl;m_1},....
          R_{ijkl;m_1....m_{p+1}}\} \ ,
\]
where $p$ is such that the components
in $R^{p+1}$ are functionally dependent on those in $R^p$ (as
functions of both the coordinates of the manifold and the parameters
of the generalized rotation group). 
This description can be used to classify geometries
\cite{Karl, AAman, MacCallumSkea} and also to construct new solutions
to Einstein's equations in terms of the set $R^{p+1}$ \cite{KarlLind, 
Bradley-Karlhede, Bradley-Marklund}. 
A set $R^{p+1}$ describes a geometry iff certain integrability
conditions are satisfied \cite{Bradley-Karlhede, Bradley-Marklund}.  
These are equivalent to the Ricci equations up to order $p+2$ and
also part of the Bianchi equations if the geometry has isometries
(note that the Ricci equations in our terminology also contain part
of the commutators). If the geometry has isometries one does not get
the full line-element with this method.
In \cite{Bradley-Marklund, Marklund} it
was shown how the method can be extended to give the basis 1-forms
${\w}^i$, and hence the full line-element in terms of canonical
coordinates, by using the isometry algebra, which is easily obtained
from $R^{p+1}$ \cite{KarlMac}.  

In an earlier paper \cite{Marklund} one of the authors used this
description to construct locally rotationally symmetric (LRS) perfect 
fluid solutions. The LRS classes I (stationary solutions with
vorticity) and III (space-homogeneous with spatial twist)
where studied. The full line-element could be constructed from
$R^{p+1}$ in all cases. In this paper we study the remaining class 
II containing solutions that in general depend on both a time-like and
a space-like coordinate, but have no vorticity or spatial twist.
There are subclasses of static solutions and spatially homogeneous
solutions in this class. The class have of course been extensively
studied \cite{vanElst-Ellis,Ellis,StewEllis,BogoNovi,Bogo,Uggla,%
  Mac2,Collins1,McVittie,Knutsen1,Knutsen2,Knutsen3,Knutsen4,%
  Knutsen5,Bonnor-Knutsen,Exact} and is perhaps the
most physically interesting among the LRS models since it contains,
e.g., solutions describing the gravitational collapse of protostars
(as far as energy dissipation may be neglected) and relativistic
spherical stars in equilibrium (if once again energy transfer is
neglected). In earlier work one has often assumed spherical
symmetry. It is worth pointing out that to many such solutions there
are corresponding solutions with hyperbolic symmetry (on the
two-dimensional surfaces of symmetry).

The set $R^{p+1}$ can be obtained from the smaller set
$S=\{R^i\!_{jkl}, \Gamma^a\!_{bi}, x^{\alpha}\!_{\vert i}\}$, where 
$\Gamma^a\!_{bi}$ ($^a\!_b = 1, .., n(n - 1)/2 - $ dim(isotropy
group)) are the part of the Ricci rotation coefficients that 
generate the rotations that do not leave $R^{p+1}$ invariant,
$x^{\alpha}$ ($\alpha = 1, ..., n - $ dim(orbits)) are essential
coordinates on the manifold (i.e., 
the elements in $R^p$ are functions only of $x^{\alpha}$ when the
frame is fixed) and the coordinate gradients are given by
$x^{\alpha}\!_{\vert i}\deq
{\bfe}_i(x^{\alpha})$ where ${\bfe}_i$ is the dual basis to
${\w}^i$. Our method have similarities with the covariant approach
by van Elst and Ellis \cite{vanElst-Ellis}, but they use the smaller
set $S'=\{R^i\!_{jkl}, \Gamma^a\!_{bi}\}$. In section 3 we give an
example of two geometries with the same set $S'$, but with different
$x^{\alpha}\!_{\vert i}$, resulting in different second covariant
derivatives of the Riemann tensor. Hence the geometries are
inequivalent and the set $S'$ is not sufficient in general to give a 
complete local description of the geometry. In this section we also
discuss the general structure of the equations and the possible
choices of coordinates. 
In section 4 the static case is treated and the set of equations is
(given an equation of state $\mu = \mu(p)$) reduced to one
integro-differential equation          
for the pressure gradient (the Tolman--Oppenheimer--Volkoff
equation). Spatially homogeneous models are studied in section 5. In
general the system is reduced to two differential equations. Tilted
models are obtained when the Gaussian curvature $K$ of the 2-surfaces 
of symmetry $\{{\der}t = {\der}r = 0\}$ equals zero. In section 6,
models depending on both a time-like and a space-like coordinate are
treated. Some probably new solutions with shear, expansion and
acceleration are given. One class is characterized by
$\sigma=-{\tfrac 16}\Theta$ (where $\sigma$ is shear and $\Theta$
expansion). A class of temporally self-similar solutions with $K=0$
and equation
of state given by $p=(\gamma-1)\mu$, where $1<\gamma<2$, is also
presented. Finally, in section 7 the full line-elements are determined
with the help of the isometry algebra.

\section{Preliminaries}

Perfect fluids with locally rotationally symmetry (LRS) may be divided
into three separate classes \cite{Marklund,vanElst-Ellis,StewEllis}.
In \cite{Marklund} one
of us studied the classes I and III. Here we treat the remaining class
II. The vorticity of the fluid, the spatial twist of the preferred
spatial direction and the magnetic part of the Weyl tensor all equal
zero in this class. 

The energy-momentum tensor for a perfect fluid is given by
${\bfT} = ({\mu} + p){\bfu}\otimes{\bfu} - p{\met} \ ,$
where ${\bfu}$ is the 4-velocity of the fluid, ${\mu}$ the energy
density, $p$ the pressure and $\met$ the metric.
We choose a comoving Lorentz-tetrad, {\ie}\ we select the tangent
basis $\{{\bfe}_i\}$ and its dual $\{ {\w}^i \}$, $i=0,...,3$, such
that ${\met} = \eta_{ij}{\w}^i\otimes{\w}^j = ({\w}^0)^2 - ({\w}^1)^2
- ({\w}^2)^2 - ({\w}^3)^2$, where ${\w}^0 = {\bfu}$.
The axes are rotated so that the LRS- symmetry lies in the 23-plane,
which makes our tetrad fixed up to rotations in that plane.

As stated above, $R^{p+1}$ can be obtained from the set
$S=\{R^i\!_{jkl}, \Gamma^a\!_{bi}, x^{\alpha}\!_{\vert i}\}$. Since
the geometry is assumed to be invariant under rotations in the 23-plane,
the $\Gamma^2\!_{3i}$ are excluded from $S$. For the same reason there
are at most two functionally independent elements in $R^{p+1}$ and all
elements can hence be seen as functions of two coordinates, say $t$
and $r$ where $t$ is time-like and $r$ is space-like. Their exterior 
derivatives can be written 
\begin{eqnarray*}
  {\der} t = X{\w}^{0} + Y{\w}^1 , \\
  {\der} r = x{\w}^{0} + y{\w}^1 , 
\end{eqnarray*}
so that the tangent basis vectors ${\bfe}_{0}$ and ${\bfe}_1$ become 
\begin{eqnarray*}
  {\bfe}_{0} = X\partial_t + x\partial_r \ , \\
  {\bfe}_1 = Y\partial_t + y\partial_r \ ,
\end{eqnarray*}
from which the coordinate gradients $x^{\alpha}\!_{\vert i}$ easily
can be read off.
The requirement for the coordinates to be independent is
$Xy - Yx \neq 0$. $t$ is time-like when $X^2 > Y^2$, and $r$ is
space-like when $y^2 < x^2$.

The Riemann tensor for a LRS perfect fluid with vanishing magnetic
part of the Weyl tensor is given by
$$
 \begin{array}{cll}
   &R_{0101} =
   E-{\tfrac16}({\mu} + 3p) \ ,&
   R_{0202} =
   R_{0303} = -\tfrac12E - {\tfrac16}({\mu} + 3p) \ , \\
   & R_{1212} = R_{1313} =  {\tfrac12}E - {\tfrac13}\mu \ ,
   & R_{2323} =  -E - {\tfrac13}\mu \ ,      
 \end{array}
$$     
where $E = E_{11}$ and 
$E_{ij} {\deq} C_{ikjl}u^{k}u^{l} = C_{0i0j}$  
denotes the electric part of the Weyl tensor.

Some of the rotation-coefficients are expressible in the kinematic
quantities acceleration $a_{i}$,
expansion $\Theta$, vorticity $\omega_{ij}$ (which is zero in our
case) and shear $\sigma_{ij}$ through 
\[
  -\Gamma_{0ij} =
  \nabla_j u_i
  =\omega_{ij}+\sigma_{ij}-{\tfrac13}h_{ij}\Theta+a_{i}u_{j} , 
\]
where \cite{Ehlers}
\[ 
 \begin{array}{cll}
   &\omega_{ij} {\deq}
   {h_i}^k{h_j}^l\nabla_{[l}u_{k]} , \quad&
 \sigma_{ij}  {\deq} 
 {h_i}^k{h_j}^l\left(\nabla_{(l}u_{k)} +
   {\tfrac13}h_{kl}\Theta \right), \\ 
 &\Theta {\deq} \nabla_iu^i  , &
 a_i {\deq} u^j\nabla_j u_i  ,
\end{array}
\]
and $h_{ij} {\deq} u_{i}u_{j} - {\eta}_{ij}$ is the projection tensor.
The choice of a comoving frame results in $a_0 =\sigma_{i0} = 0$,
because of their orthogonality to the fluid 4-velocity.
From the requirement of LRS there is only one component of the
acceleration, $a_1 {\deq} a$. The shear will, because it is traceless,
look like $(\sigma_{ij}) = \rm{diag}(0, -2\sigma, \sigma, \sigma)$. 
The necessary non-zero rotation coefficients for LRS class II are then
\[
\begin{array}{cll}
  &\Gamma_{010} = -a\ , 
  &\Gamma_{011} =
  2\sigma + {\tfrac13}\Theta \ \deq {\al} , \\ 
  &\Gamma_{022} = \Gamma_{033}
  =  -\sigma +
      {\tfrac13}\Theta \ \deq {\be},   \quad
  & \Gamma_{122} = \Gamma_{133} {\deq}  -\kappa\ ,    
\end{array}
\]
(see \cite{Marklund} for the general LRS case).
The rotation coefficient $\kappa$ corresponds to the spatial
divergence of the vector-field ${\bfe}_1$. Thus all rotation
coefficients are covariantly defined because of the LRS, and we have
the following set of functions determining the geometry in the generic
case: 
\[
  S = \{E, \mu, p, a, \alpha, \beta, \kappa, X, Y, x, y, \eta_{ij}\} \
  . 
\]

We would like to stress the fact that the variables we have chosen to
describe the geometry are only singled out because of their direct
correspondence with the components of the Riemann tensor and the Ricci
rotation coefficients. Any other set that can reproduce the set
$R^{p+1}$ could be used as well.

The functions in the set $S$ are subject to the following 
integrability conditions:
\begin{eqnarray}
  {x^{\alpha}}_{|[k,|{\beta}|}{x^{{\beta}}}_{|l]} =
  {x^{\alpha}}_{|m} {{\Gamma}^{m}}_{[kl]}\ ,\label{eq:Comm}\\
  {R^{a}}_{bij} = 2{{\Gamma}^{a}}_{b[j,|\alpha |}{x^{\alpha}}_{|i]} +
  2{{\Gamma}^{am}}_{[j}{{\Gamma}}_{|bm|i]} +
  2{{\Gamma}^{a}}_{bk}{{\Gamma}^{k}}_{[ij]} \ , \label{eq:RieEq} \\
  R^t\!_{[ijk]} = 0 \ , \label{eq:Cycl}\\
  R^p\!_{q[ij;k]} = 0\ . \label{eq:Bian}
\end{eqnarray} 
Here $\alpha, \beta = 1, 2, ..., l$, $\{^a\!_b\} = 1, 2, ..., m$,
$t=l + 1, l + 2,..., n$, and $\{^p\!_q\} = m + 1, m + 2, ..., n(n -
1)/2$, where $l := n - \rm{dim(orbits)}$ and $m := n(n-1)/2 -
\rm{dim(isotropy \,\, group)}$ (see Refs.
\cite{Bradley-Karlhede, Bradley-Marklund} for a
more detailed exposition of the method). For the case at hand they
are: 
 
\noindent\textit{Commutator equations}
\begin{eqnarray}
  \left[ {\bfe}_1,{\bfe}_{0} \right] t = {\bfe}_1(X)
  - {\bfe}_{0}(Y) =
   \dot{X}Y + X'y - \dot{Y}X - Y'x   
   =  Xa + Y{\al} \ , \label{eq:commutator1}\\  
   \left[ {\bfe}_1,{\bfe}_{0} \right] r = {\bfe}_1(x)
   - {\bfe}_{0}(y) =
   \dot{x}Y + x'y -  \dot{y}X - y'x   
   = xa + y{\al},\label{eq:commutator2}
\end{eqnarray}
 
\noindent\textit{Ricci equations}
\begin{eqnarray}
   {\bfe}_{0}({\al}) + {\bfe}_1(a) = 
   X\dot{\al} + x{\al}' + Y\dot{a} + ya' 
   =  E - {\tfrac16}(\mu + 3p) + a^{2} - {{\al}}^{2} ,
   \label{eq:ricci1} \\ 
    {\bfe}_{0}({\be}) = X\dot{{\be}} + x{{\be}}'  = 
   -{\tfrac12}E - {\tfrac16}({\mu} + 3p) + a{\kappa} - {{\be}}^{2} \ ,
   \label{eq:ricci2}\\ 
    {\bfe}_1({\be}) = Y\dot{{\be}} + y{{\be}}' =  
   {\kappa}({{\be}} - {{\al}}) , \label{eq:ricci3}\\
     {\bfe}_{0}(\kappa) = X\dot{\kappa} + x{\kappa}' =  
   (a - {\kappa}){{\be}} , \label{eq:ricci4}  \\
    {\bfe}_1(\kappa) = Y\dot{\kappa} + y{\kappa}' = 
   -{\tfrac12}E + {\tfrac13}{\mu} + {\kappa}^{2}   
   - {\al}{\be}, \label{eq:ricci5}
\end{eqnarray}
 
\noindent\textit{Bianchi equations} 
\begin{eqnarray}
  {\bfe}_{0}(E + {\tfrac13}\mu) =
  X{\left(\dot{E} + {\tfrac13}\dot{{\mu}} \right)} 
   + x{\left( E' + {\tfrac13}\mu' \right)} =  
   -(3E + {\mu} + p){{\be}} , \label{eq:bianchi1}  \\
  {\bfe}_1(E + {\tfrac13}\mu) = Y{\left( \dot{E} +
       \tfrac13\dot{{\mu}} \right)} 
  + y{\left( E' + {\tfrac13}\mu' \right)} =  3E{\kappa} \
  . \label{eq:bianchi2} 
\end{eqnarray}
Here we have used the notation $\dot{f} := \partial_tf$ and
$f' := \partial_rf$.
 
Observe that the cyclic equations ${R^{t}}_{{[}ijk{]}}   =  {0}$ are 
already satisfied by our choice of Riemann tensor.
 
It might be preferable to replace the electric part $E$ of the Weyl
tensor by the Gaussian curvature $K {\deq} E + \mu/3 + \kappa^2 -
\beta^2$ of the orbits $\{\der t = \der r = 0\}$. The Ricci identities
then become 
\begin{eqnarray}
     {\bfe}_{0}({\al}) + {\bfe}_1(a)
    = K -{\tfrac12}(\mu + p) - \kappa^2 + a^2 - \alpha^2 + \beta^2 \ ,
    \label{eq:ricci1'} \\
    {\bfe}_{0}(\beta)  = 
    - {\tfrac12}K - {\tfrac12}p + {\tfrac12}\kappa^2 + a\kappa - 
    {\tfrac32}{\be}^2 \ ,
    \label{eq:ricci2'}
    \\
    {\bfe}_1(\beta) =  
    \kappa(\beta - \alpha) ,
    \label{eq:ricci3'}
    \\
    {\bfe}_{0}(\kappa) =  
    (a - \kappa)\beta \ ,
    \label{eq:ricci4'}
    \\
    {\bfe}_1(\kappa) = 
   - {\tfrac12}K + {\tfrac12}\mu + {\tfrac32}\kappa^2 -
    {\tfrac12}\beta(2\alpha + \beta) 
   \ .
   \label{eq:ricci5'}
\end{eqnarray}
The Gaussian curvature $K$ satisfies the equations
\begin{eqnarray}
  {\bfe}_{0}(K) = X\dot{K} + xK' = -2\beta K \ ,
  \label{eq:bianchi1'}
  \\
  {\bfe}_1(K) = Y\dot{K} + yK' = 2\kappa K \ ,
  \label{eq:bianchi2'}
\end{eqnarray}
which can be used to replace the Bianchi equations.

\section{General considerations}\label{sec:GC}

In this section we discuss the choice of coordinates, the algorithm  
for solving the equations in the generic case and give an example
of where essential information about a geometry is hidden in the
coordinate gradients.

First maybe a word about the meaning of sufficient and necessary conditions
is at hand. The equations (\ref{eq:commutator1})--(\ref{eq:bianchi2})
are sufficient, i.e., a solution to them really describes a geometry,
but the consistency equation for, e.g.,
(\ref{eq:ricci2})--(\ref{eq:ricci3}) could give ``new'' equations (in
this paper one of the twice contracted Bianchi identities) that are
helpful in solving the set of equations. 

In general, there exists several ways to choose ones coordinates.
Components of the curvature tensor, the rotation coefficients, and/or
the coordinate gradients may be used as coordinates.

If we choose some of the rotation coefficients as
coordinates, this is in line with the integration procedure in
the GHP formalism \cite{GHP} suggested by Held \cite{Held1,Held2}
and developed further in several papers by Edgar \cite{Edgar1}
and Edgar and Ludwig \cite{Edgar2,Edgar3,Edgar4}.\footnote{In
  \cite{Edgar1} Edgar also pointed out the necessity of the
  commutator equations in the GHP formalism.}
The commutators will generate Bianchi \emph{type} equations, {\ie}, 
equations containing derivatives of the Riemann tensor components. In 
general there may be new equations among these (see section
\ref{sec:generic}).

\subsection{The generic case}\label{sec:generic}

The theory developed in \cite{Bradley-Karlhede,Bradley-Marklund}
now assures that a solution to the commutator 
equations (\ref{eq:commutator1})--(\ref{eq:commutator2}),
the Ricci equations (\ref{eq:ricci1})--(\ref{eq:ricci5}) (or
equivalently equations (\ref{eq:ricci1'})--(\ref{eq:ricci5'})), and
the Bianchi equations (\ref{eq:bianchi1})--(\ref{eq:bianchi2}) (or 
equivalently equations (\ref{eq:bianchi1'})--(\ref{eq:bianchi2'}))
describes a LRS perfect fluid solution to Einstein's equations.
Since some of the quantities appear differentiated in more than one
equation ($\kappa$, $\beta$ and $K$) it could be helpful to calculate
the consistency conditions that these equations give rise to. These
are 
\[
  \left[{\bfe}_1,{\bfe}_0\right]f = \left(\Gamma^k\!_{01} -
    \Gamma^k\!_{10}\right){\bfe}_k(f) = a{\bfe}_0(f) +
  \alpha{\bfe}_1(f)
\]
where $f$ is $\kappa$, $\beta$ or $K$ respectively. By acting with
${\bfe}_0$ and ${\bfe}_1$ on
(\ref{eq:ricci2})--(\ref{eq:ricci5}) one then gets
\begin{eqnarray}
  {\bfe}_0(\mu) = -\left(\alpha + 2\beta\right)(\mu + p) \ ,
  \label{eq:contrBian1} \\
  {\bfe}_1(p) = a(\mu + p) \ , \label{eq:contrBian2}
\end{eqnarray}
{\ie}, the twice contracted Bianchi equations (note that these are not
contained in equations (\ref{eq:bianchi1'})--(\ref{eq:bianchi2'})),
whereas the commutator relation acting on $K$ is identically
satisfied.  

Let us first look at the \textit{generic} case when two functionally
independent quantities can be found among $\kappa$, $\beta$ and
$K$. As an example we choose $\beta$ and $\kappa$ as the coordinates 
$t$ and $r$. Equations (\ref{eq:ricci2'})--(\ref{eq:ricci5'}) then
give 
\begin{eqnarray*}
  X = -\tfrac12K - \tfrac12p + \tfrac12\kappa^2 + a\kappa -
  \tfrac32\beta^2 \ , \\
  Y = \kappa\left(\beta - \alpha\right) \ , \\
  x = (a - \kappa)\beta \ , \\
  y = -\tfrac12K + \tfrac12\mu + \tfrac32\kappa^2 -
  \tfrac12\beta\left(2\alpha + \beta\right) \ . 
\end{eqnarray*}

This choice can be made as long as $Xy-Yx \neq 0$. The commutator
equations (\ref{eq:commutator1})--(\ref{eq:commutator2}) are now
identically satisfied due to (\ref{eq:ricci1'})--(\ref{eq:ricci5'})
and (\ref{eq:contrBian1})--(\ref{eq:contrBian2}) (naturally, since the
commutators on $t$ and $r$ now coincide with the commutators on
$\beta$ and $\kappa$). We hence may use the twice contracted Bianchi
equations (\ref{eq:contrBian1})--(\ref{eq:contrBian2}) instead of the
commutators (\ref{eq:commutator1})--(\ref{eq:commutator2}) and we get
the same set of equations as in \cite{vanElst-Ellis} (our set of
equations (\ref{eq:ricci1})--(\ref{eq:bianchi2}) and
(\ref{eq:contrBian1})--(\ref{eq:contrBian2})). 

Using equations (\ref{eq:ricci2'})--(\ref{eq:ricci5'}) for defining
$X,x,Y$ and $y$ there now remains five differential equations to
solve, (\ref{eq:ricci1'}), (\ref{eq:bianchi1'})--(\ref{eq:bianchi2'}),
and (\ref{eq:contrBian1})--(\ref{eq:contrBian2}).
From (\ref{eq:ricci1'}) we see that one of $a$ and $\alpha$, say $a$,
may be freely specified. The system is then integrable since the
commutator acting on $K$ is identically satisfied and we just have one
equation for each of $\alpha$, $\mu$ and $p$. More natural would of
course be to specify an equation of state $p=p(\mu)$. Instead of
equation (\ref{eq:contrBian2}) we then get  
\begin{equation}
  {\bfe}_1(\mu) = a(\mu + p)\left(\frac{dp}{d\mu}\right)^{-1} \ .
  \label{eq:contrBian2'}
\end{equation}

The consistency condition for equations (\ref{eq:contrBian1}) and
(\ref{eq:contrBian2'}) (as equations for $\mu$) is then
\begin{eqnarray}
  \fl {\bfe}_1(\alpha ) + {\bfe}_0(a)\left(\frac{dp}{d\mu}\right)^{-1}
  & = &2\kappa(\alpha - \beta)
  - a\alpha\left(\frac{dp}{d\mu}\right)^{-1} \nonumber \\
  &&\quad + a(\alpha + 2\beta) - a(\alpha + 2\beta)(\mu + p) 
  \frac{d^2p}{d\mu^2}\left(\frac{dp}{d\mu}\right)^{-2} \
  . \label{eq:ricci1''} 
\end{eqnarray}   
Equations (\ref{eq:ricci1'}) and (\ref{eq:ricci1''})
may be considered as developing equations 
along ${\bfe}_0$ for $\alpha$ and $a$ respectively. They may then be 
specified on a three dimensional space-like hypersurface 
(since
$\alpha$ and $a$ are specified on this surface, their gradients along
${\bfe}_1$, which is tangent to the surface, may be calculated) (cf.\
e.g.\ \cite{Petrovsky}). We
hence have the following problem to solve (see figure 1):
Given an equation of state
$p=p(\mu)$, the development of $\mu$, $K$, $\alpha$
and $a$ is given by  (\ref{eq:contrBian1}) and (\ref{eq:contrBian2'}),
(\ref{eq:bianchi1'}) and (\ref{eq:bianchi2'}), and (\ref{eq:ricci1'}) 
and (\ref{eq:ricci1''}) respectively, with $\alpha$ and $a$
specified on a space-like hypersurface orthogonal to ${\bfe}_0$ and
$\mu$ and $K$ specified on a two-dimensional orbit contained in the
hypersurface. Note that all these considerations are purely local,
since in a general space-time a foliation with space-like
hypersurfaces may not be possible. 
\begin{figure}
  \centering
  {\epsfig{figure=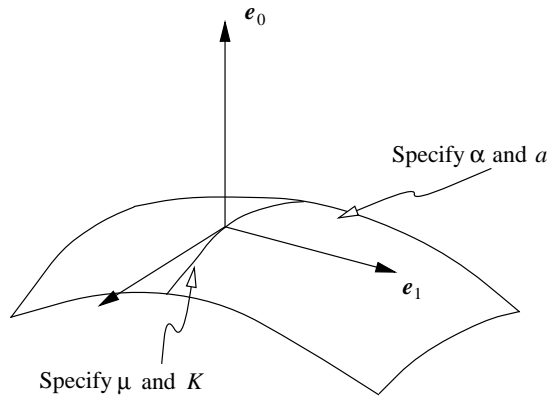}}
  \caption{Boundary value specification when $p = p(\mu)$.}
\end{figure}

\subsection{A geometry that is not completely determined by the
the Riemann tensor and the rotation coefficients}

In the example above the commutators
(\ref{eq:commutator1})--(\ref{eq:commutator2}) could be neglected when
the twice contracted Bianchi equations where added, but cases when
this is \emph{not} possible do appear. Assume that all the Ricci rotation
coefficients and all the components of the Riemann tensor only depend
on one functionally independent quantity (which is chosen as the first
coordinate). Nevertheless the geometry could be truly depending on two 
coordinates, the second one appearing first in the first covariant
derivative of the Riemann tensor. It must then be found in one or both
of the vectors ${\bfe}_0$ and ${\bfe}_1$, since the
first derivative should be constructible from the vectors, the 
Riemann tensor and the rotation coefficients. We here give an example
of such a metric. Assume that $K=\kappa=a=0$, that $\beta$ and $p=-3
\beta^2$ are constants and that $\mu$ is a function of $\alpha$
(chosen to be the first coordinate). $X$ is then found to be $X =
2\beta^2 - \alpha^2 - \alpha\beta$. Even if we add all of the Bianchi
equations, we will not get any more information, and we still have no 
equations for $Y$, $x$ or $y$. But choosing them completely freely
would contradict the commutators
(\ref{eq:commutator1})--(\ref{eq:commutator2}) and we hence cannot
ignore these equations. If we choose $Y$ as our second coordinate,
$x=-(3 \alpha + \beta)Y$ from equation (\ref{eq:commutator1}), and
(\ref{eq:commutator2}) gives a differential equation for $y$:
\[
-3Y^2 - (3\alpha + \beta)y-\dot{y}\left(2\beta^2 - \alpha^2 -
\alpha\beta\right) + y'(3\alpha + \beta)Y = y\alpha \ ,
\]
with solution $y = (C - 3\alpha)Y^2/\left(2\beta^2 - \alpha^2 -
\alpha\beta\right)$, where $C$ is
a constant of integration. The coordinates $\alpha$ and $Y$ will be
linearly independent as long as $Xy-xY \neq 0$, corresponding to $C + 
\beta \neq 0$. If one now calculates the set $R^{p+1}$ one finds the
second essential coordinate $Y$ in the first covariant derivative
({\eg}\ $R_{0101;1} = -\beta Y$) showing that the metric really is
depending on two essential coordinates. Furthermore, the constant $C$
appears first in the second covariant derivative ({\eg}\ $R_{0101;11}
= -\alpha^3\beta - \alpha^2\beta^2 + 3\alpha\beta Y^2/X - C\beta
Y^2/X$) showing that metrics with different $C$ are geometrically
inequivalent, and that the second derivative is needed to see this.

The 1-forms for this metric are given by (see also section
\ref{sec:Isometry}):
\begin{eqnarray*}
  {\w}^0 = \frac{C - 3\alpha}{(C+ \beta)\left(2\beta^2 - \alpha^2 -
    \alpha\beta\right)}{\der}\alpha - \frac{1}{(C+\beta)Y}{\der}Y \ ,
    \\ 
  {\w}^1 = \frac{3\alpha + \beta}{(C + \beta)Y}{\der}\alpha 
  + \frac{2\beta^2 - \alpha^2 - \alpha \beta}{(C+\beta)Y^2}{\der}Y
  \ , \\ \bs
  {\w}^2 = |Y|^{-\beta/(C + \beta)}\left(|\alpha -
    \beta|\right)^{(3\beta - C)/[3(\beta + C)]}\left(|\alpha +
    2\beta|\right)^{(6\beta + C)/[3(\beta + C)]}{\der}u \ , \\
  {\w}^3 = |Y|^{-\beta/(C + \beta)}\left(|\alpha -
    \beta|\right)^{(3\beta - C)/[3(\beta + C)]}\left(|\alpha +
    2\beta|\right)^{(6\beta + C)/[3(\beta + C)]}{\der}v \ .
\end{eqnarray*}

\section{Static models}\label{sec:static}

In this case we have no dependence on the coordinate
$t$, thus we have $X = Y = 0$. The functions $\beta$, $\kappa$ and $K$
then have two differential equations each, so that we may obtain
algebraic relations for some of these.

\subsection{Static models with $K \neq 0$}

If $K \neq 0$ we may solve for $\beta$ as $\beta = -x\kappa/y$ from
equations (\ref{eq:bianchi1'}) and (\ref{eq:bianchi2'}). Inserting
this definition into equation (\ref{eq:ricci2'}), we get an algebraic
expression for, say, $K$. Combining equations (\ref{eq:ricci4'}) and
(\ref{eq:ricci5'}) and substituting the algebraic equation for $K$, we
find $x(\mu + p) = 0$, i.e., $x = 0$
With $x = 0$, the commutator (\ref{eq:commutator2})gives $\alpha = 0$,
and from the Bianchi equation (\ref{eq:bianchi1'}) 
we see that $\beta = 0$. Thus the static models with non-planar
symmetry are expansion and shear free. From (\ref{eq:ricci2'}) we have
\begin{equation}\label{eq:algconstr}
  K=-p+\kappa^2+2a\kappa \ ,
\end{equation}
and the remaining differential equations are
\begin{eqnarray}
  ya'      = 2a\kappa + a^2 - {\tfrac12}(\mu+3p) \ ,
  \label{eq:stat1} \\
  y\kappa' = -a\kappa + \kappa^2 + {\tfrac12}(\mu+ p) \ ,
  \label{eq:stat2} \\
  yp'      = a(\mu+p) \ . \label{eq:stat3}
\end{eqnarray}
Since we can choose one of the functions as our variable $r$, we see  
that there remains to solve two coupled, ordinary differential
equation.

When we have \emph{constant pressure}, equation (\ref{eq:stat3}) gives
$a = 0$, and equation (\ref{eq:stat1}) gives $\mu + 3p = 0$. The
only remaining equation is (\ref{eq:stat2}). If $\kappa$ is constant
we have a space-time homogeneous case, and if $\kappa$ is non-constant
we may choose it as our coordinate, which implies $y = \kappa^2 - p$. 

When there is a \emph{non-constant pressure}, it is natural to take
the pressure as our coordinate. From equation (\ref{eq:stat3}) we find
$y = a(\mu + p)$. The remaining equations are
\begin{eqnarray*}
  (\mu + p)aa'      = a(2\kappa + a)
  - {\tfrac12}(\mu + 3p) \  , \\
  (\mu + p)a\kappa' = \kappa(\kappa - a)
  + {\tfrac12}(\mu + p) \ .
\end{eqnarray*}
Most constraints on these equations will reduce the problem to an
algebraic one.
For instance, if $a = \kappa$, we find that $\mu = p + A$, where $A$ 
is an integration constant, and $\kappa^2 = \left(3p +
  2A\right)/3$.

We may of course use another set of equations than the one chosen
above. Instead of eliminating $K$ with the help of equation
(\ref{eq:algconstr}), we may focus on eliminating $a$ as
\begin{equation}\label{eq:asolve}
  a = \frac{K + p - \kappa^2}{2\kappa} \ .
\end{equation}
This is especially useful when treating spherically
symmetric space-times, i.e., $K > 0$. We define our radial coordinate
$r$ through $K \deq r^{-2}$, so that from equation
(\ref{eq:stat1}) we have
\[
  y = -r\kappa \ .
\]
Using equation (\ref{eq:algconstr}) to eliminate $p$ we find that
equation (\ref{eq:stat2}) becomes
\[
  \frac{d\kappa^2}{dr} + \frac{3\kappa^2}{r} = \frac{1-\mu r^2}{r^3} 
  \ ,
\]
which may be integrated in terms of $r$ and $\mu$:
\[
  \kappa^2(r) = \frac{4\pi r - m}{4\pi r^3} \ ,
\]
where we have introduced the mass function
\begin{equation}\label{eq:massa}
  m(r) \deq 4\pi\int_0^r\mu(\hat{r})\hat{r}^2\der\hat{r} \ ,
\end{equation}
and assumed that there are no point masses at the center ($r = 0$) of
the model. Eliminating $a$ from equation (\ref{eq:stat3}) by using
equation (\ref{eq:asolve}), we obtain
\begin{equation}\label{eq:TOV}
  \frac{dp}{dr} = -\frac{(\mu + p)(4\pi pr^3 + m)}{2r(4\pi r - m)} \ ,
\end{equation}
which is the Tolman--Oppenheimer--Volkoff equation
\cite{Zeldovich-Novikov}. Hence, given an equation of state
$p=p(\mu)$, one needs to solve (\ref{eq:TOV}) together with the
definition (\ref{eq:massa}) to get the full solution.
In the usual approach with a metric ansatz there will remain
one further equation to be
integrated for obtaining the full line-element.
Here we see that this is not
necessary for a complete local description of the space-time.

We may also take another route in analyzing the equations for static
LRS class II space-times. Since we only have dependence on one
variable, and we only have one coordinate gradient, we may view $y$ as
a lapse function. Thus by defining $\der r \deq y\der R$,  
equations (\ref{eq:stat1})--(\ref{eq:stat3}) becomes an
autonomous system of equations:
\begin{eqnarray*}
  \frac{da}{dR}      = 2a\kappa + a^2 - {\tfrac12}(\mu+3p) \ ,
  \frac{d\kappa}{dR} = -a\kappa + \kappa^2 + {\tfrac12}(\mu+ p) \ , 
  \frac{dp}{dR}      = a(\mu+p) \ .
\end{eqnarray*}
This set of equations is the starting point for a dynamical systems 
analysis, although a change of variables is in place to acquire
dimensionless quantities and a compact phase space.

\subsection{Static models with $K = 0$}
When $K = 0$, the Bianchi equations can no longer be used to reduce 
the set of variables and equations. Instead, we will have
\begin{eqnarray}
  yx' - xy' = xa + y\alpha \, \label{eq:statcom} \\
  x\alpha' + ya' = -\tfrac12(\mu + p) - \kappa^2 + a^2 - \alpha^2
  + \beta^2 \ , \label{eq:statric1} \\
  x\beta' = -\tfrac12p + \tfrac12\kappa + a\kappa -\tfrac32\beta^2 \
  , \label{eq:statric2} \\
  y\beta' = \kappa(\beta - \alpha) \ , \label{eq:statric3} \\
  x\kappa' = (a - \kappa)\beta \ , \label{eq:statric4} \\
  y\kappa' = \tfrac12\mu + \tfrac32\kappa^2 - \alpha\beta -
  \tfrac12\beta^2 \ . \label{eq:statric5}
\end{eqnarray}

Now, if $x = 0$, the commutator equation (\ref{eq:statcom})
becomes $\alpha = 0$. Equation (\ref{eq:statric4}) gives (i) $a =
\kappa$, or (ii) $\beta = 0$. Case (i) gives $\mu + p = 0$, which is
equivalent to vacuum with a cosmological constant. Case (ii) results
in 
\begin{eqnarray}
  ya' = -\tfrac12\mu - \tfrac32\kappa^2 - a\kappa + a^2 \ ,
  \label{eq:XX} \\
  y\kappa' = \tfrac12\mu + \tfrac32\kappa^2 \ ,
\end{eqnarray}
and
\[
  p = \kappa^2 + 2a\kappa \ .
\]
If we choose $\kappa = r$ $\Rightarrow$  $y = \mu/2 +3\kappa^2/2$ and 
equation (\ref{eq:XX}) becomes an equation for $\mu$, while $a$ is an 
arbitrary function.

If $x \neq 0$, i.e., the metric is manifestly non-diagonal, we can
replace equations (\ref{eq:statric2}) and (\ref{eq:statric5}) with two
algebraic equations for $p$ and $\mu$:
\begin{eqnarray*}
  p = \kappa^2 + 2a\kappa - 3\beta^2 - \frac{2x(\beta -
  \alpha)\kappa}{y} \ , \\
  \mu = -3\kappa^2 + \beta(2\alpha + \beta) + \frac{2y(a -
    \kappa)\beta}{x} \ .
\end{eqnarray*}
The remaining equations are
\begin{eqnarray*}
  y\beta' = (\beta - \alpha)\kappa \ , \\
  x\kappa' = (a - \kappa)\beta \ , \\
  x\alpha' + ya' = -\frac{y(a - \kappa)\beta}{x} + \frac{x(\beta -
    \alpha)\kappa}{y} \nonumber \\
  \qquad\qquad\quad  - a\kappa + a^2 - \alpha^2 + 2\beta^2 -
  \alpha\beta \ , \\
  x'y - y'x = xa + y\alpha \ .
\end{eqnarray*}
There are in total 6 functions. Choosing one of the functions as
coordinate, we determine $x$ or $y$. Then there are 4 functions 
and 3 equations left, thus one of the functions is arbitrary
(as in the case $x = 0$). 

\section{Spatially homogeneous models}\label{sec:homo}
Here we have no dependence on the coordinate $r$. This
will, as in the static case, give some algebraic equations from which 
we may express suitable quantities.

\subsection{Spatially homogeneous models with $K \neq 0$}

When $K \neq 0$, we use the Bianchi equations (\ref{eq:bianchi1'}) and
(\ref{eq:bianchi2'}) to obtain $\kappa = -Y\beta/X$. From 
equation (\ref{eq:ricci5'}) we then obtain an algebraic equation for
$K$. Using equations (\ref{eq:ricci2'}) and (\ref{eq:ricci3'}) is is
straightforward to show that $Y(\mu + p) = 0$, i.e., $Y = 0$.

The curvature of the 2-surfaces $\{t = \mathrm{constant}, r =
\mathrm{constant}\}$ is related to the scalar curvature ${^{(3)}R}$ of
the 3-surfaces of homogeneity through the Gauss--Codazzi
equation. Using this, we find that 
\[
  {^{(3)}R} = 2K \ .
\] 

If $Y = 0$, the commutator (\ref{eq:commutator1}) gives $a = 0$, and
from the Bianchi equation (\ref{eq:bianchi2'}) we get $\kappa = 0$,
while (\ref{eq:ricci5'}) becomes
\begin{equation}
  \mu = K + 2\al\be + \be^2 \ . \label{eq:alg}
\end{equation}
The remaining equations are
\begin{eqnarray}
  X\dot{\al} = {\tfrac12}K - \al\be - \al^2 + {\tfrac12}\be^2 -
  {\tfrac12}p \ ,
  \label{eq:homogen1} \\
  X\dot{\be} = -{\tfrac12}K - {\tfrac32}\be^2 - {\tfrac12}p \ ,
  \label{eq:homogen2} \\
  X\dot{K} = -2\be K \ .\label{eq:homogen3}
\end{eqnarray}
It turns out this system of equations can be reduced to one Riccati
equation. First we note that if $K$ is a non-zero constant,
equations (\ref{eq:homogen1})--(\ref{eq:homogen3}) gives $\beta =
0$, $p = -K$ $\Rightarrow$ $\mu + p = 0$ by (\ref{eq:alg}). 
If $\dot{K} \neq 0$,
we can choose $K$ as coordinate (so
that $X = -2\be K$ by (\ref{eq:homogen3})), and
equation (\ref{eq:homogen2}) becomes linear in $\be^2$. Further, if we
assume that $p = p(K)$ is a given function, this linear equation may
be integrated:
\[
  \beta^2(K) = CK^{3/2} - K +
  {\tfrac12}K^{3/2}\int^{K}p(\hat{K})\hat{K}^{-5/2} \der\hat{K} \ .
\]
Thus the equation for $\al$ becomes
\begin{equation}
  \frac{d\al}{dK} = F(K)G(K) + \be G(K)\al + G(K)\al^2 \ ,
  \label{eq:riccati} 
\end{equation}
where $F(K) \deq (p - \be^2 - K)/2$ and $G(K) \deq 1/(2\be K)$. The
Riccati equation (\ref{eq:riccati}) can be solved exactly when, e.g.,
there is some algebraic relationship between the coefficients $F$ and
$G$.
Such a relation will in general lead to an integral equation for
the function $p$ in terms of $K$. Thus, although the choice of
variables leads us to a single equation for these space-times, the
structure of the remaining equation leads to difficulties concerning 
the equation of state.

A perhaps physically more suitable choice of variables is $\Theta$
(expansion), $\sigma$ (shear), and $\mu$ (where we have used that
$\mu$ and $K$ are interchangeable, by the algebraic equation
(\ref{eq:alg})), which leads to
\begin{eqnarray}
  X\dot{\Theta} =  - {\tfrac13}\Theta^2 - 6\sigma^2  -{\tfrac12}(\mu
  + 3p) \ ,
  \label{eq:homogen1'} \\
  X\dot{\sigma} = - {\tfrac13}\Theta^2 + 3\sigma^2 - 3\Theta\sigma +  
  \mu \ , 
  \label{eq:homogen2'} \\
  X\dot{\mu} = -\Theta(\mu + p) \ .
  \label{eq:homogen3'} 
\end{eqnarray}
Assuming non-constant density, we can solve equations
(\ref{eq:homogen1'}) and (\ref{eq:homogen2'}) for $\Theta$ and
$\sigma$, respectively, in
terms of $\mu$, if we have a barotropic equation of state.

We can look at the interesting case when the shear is small, i.e.,
when the deviations from isotropy are small. This translates into
$\sigma/\Theta \ll 1$. Linearizing the equations with respect to this
parameter, we can solve for the expansion and shear in terms of the
density:
\begin{eqnarray}
  \Theta^2 = F^{2/3}(\mu)\left[
  \int^{\mu}F^{-2/3}(\hat\mu)\frac{\hat\mu + 3p(\hat\mu)}{\hat\mu +
    p(\hat\mu)}\der\hat\mu + A \right] \ , \label{eq:expansion} \\
  \sigma = F(\mu)\left[ \int^{\mu}
    F^{-1}(\hat\mu)\frac{\Theta^2(\hat\mu) -
  3\hat\mu}{9\Theta(\hat\mu)(\hat\mu + p(\hat\mu))}\der\hat\mu +
   B\right] \ , \nonumber
\end{eqnarray}
where $A$ and $B$ are constants, and $F(\mu) \deq
\exp\left[\int^{\mu}(\hat\mu + p(\hat\mu))^{-1}\der\hat\mu\right]$. 

\subsection{Friedmann--Lema{\^i}tre--Robertson--Walker models}
These space-times are characterized by the existence of a
3-dimensional spatial isotropy group and a 6-dimensional isometry
group acting multiply transitively on 3-dimensional orbits. 
Therefore $E = a = \sigma = Y = 0$, and $\kappa$ 
is a redundant variable. (Since $\kappa$ is no longer part of
  the set $S$, it is possible for it to depend on some spatial
  variable).
This of course means that there is need for additional Bianchi
equations, and we need fewer Ricci equations. It turns out that the
additional equations are just the twice contracted Bianchi
equation (\ref{eq:contrBian1}). The redundant
Ricci equations are the ones containing $\kappa$. Thus,
(\ref{eq:homogen1}) and (\ref{eq:homogen3})
are the necessary and sufficient equations. 
Choosing $\mu$ as coordinate and an equation of state $p =p(\mu)$, 
the system is easily integrated to give equation (\ref{eq:expansion}). 

\subsection{Homogeneous solutions with $K=0$}\label{sec:Flat}

When $K=0$ it is possible to have tilted models, i.e., models where
the matter flow is not orthogonal to the surfaces of homogeneity. This
means that $Y\neq0$.

The situation is similar to the plane symmetric static case. In fact,
the set of equations becomes the same as equations
(\ref{eq:statcom})--(\ref{eq:statric5}), with the replacements $x, y
\rightarrow X, Y$ and $({\partial}/{\partial}r) \rightarrow
({\partial}/{\partial}t)$. For the generic case with $Y \neq 0$ the
treatment is identical to the one in the static case. For $Y = 0$, one
obtains $a = 0$ and $\kappa(\beta - \alpha) = 0$, i.e., $\kappa = 0$
or $\alpha = \beta$. In both cases the systems can be completely
integrated, similarly to the static case.

If one assumes dust, i.e., $p=0$, the equations
may be completely integrated as was shown already in
\cite{Farnsworth}.
If $\kappa$ is used as time coordinate, one gets
\[
  X=\mp \kappa^2 \sqrt{1+C\kappa} \ , \ \   
  Y=\kappa^2-\frac{\mu}{3C\kappa} \ , \ \
  \beta=\pm\kappa\sqrt{1+C\kappa}  \ ,
\]
\[
  \alpha=\mp\frac{C\kappa^2}{2\sqrt{1+C\kappa}}\pm
    \frac{\mu(3C\kappa+2)}{6C\kappa^2\sqrt{1+C\kappa}} \ , \ \
    \mu=\frac{3CD\kappa^3}{D-\kappa\sqrt{1+C\kappa}} \ ,
\]
where $C$ and $D$ are constants of integration. There is also an
isolated solution $\mu=3C\kappa^3$, but then $Y=0$. If $\kappa$
cannot be used as coordinate, i.e., if it equals a constant, it
also follows that $Y=0$.

\section{Inhomogeneous, time dependent models}

Herein lies perhaps the most interesting models, such as 
collapsing perfect fluids with an inhomogeneous matter
distribution. Since no special simplifications appear in this
class, the system of equations is in general impossible to solve
explicitly. One therefore has to resort to simplifying assumptions.
Some special cases are treated below.

No particular simplification is implied by the constraint $K=0$, since
the Bianchi equations are identically satisfied by this choice. The
analysis of the system of equations characterizing these space-times 
can be performed analogously to the case $K \neq 0$.

Before turning to the examples we show how our set of equations may be
written in a form similar to that one would obtain from a metric ansatz.
In the generic case with $K \neq 0$, we may proceed as follows:
Define $a$ and $\alpha$ through equations (\ref{eq:commutator1}) and
(\ref{eq:commutator2}), thus expressing them as derivatives of the
coordinate gradients. Further, define $\beta$ and $\kappa$ by
equations (\ref{eq:bianchi1'}) and (\ref{eq:bianchi2'}), respectively.
From equations (\ref{eq:ricci2'}) and (\ref{eq:ricci5'}) we find $p$
and $\mu$ respectively. We are then left with the second order
equations (\ref{eq:ricci1'}), (\ref{eq:ricci3'}), and
(\ref{eq:ricci4'}) for the variables $K$, $x$, $y$, $X$, and $Y$.
It is straightforward to show that equations (\ref{eq:ricci3'}) and
(\ref{eq:ricci4'}) are the same.
By performing a coordinate transformation $t' = t'(t,r)$, $r' =
r'(t,r)$ it is always possible to get $Y = 0$ and $x = 0$, hence
making $\partial/\partial t'$ parallel to the four-velocity of the
fluid. The remaining freedom in choice of coordinates are now
transformations of the type $r'' = r''(r')$, $t'' = t''(t')$. 
With this choice of coordinates, equations (\ref{eq:ricci3'}) and 
(\ref{eq:ricci1'}) reduce to
(dropping all the primes) 
\begin{eqnarray}
  \fl \frac{\partial^2\xi}{\partial t\partial r}
  + \frac{\partial\chi}{\partial r}\frac{\partial\xi}{\partial t}
  + \frac{\partial\upsilon}{\partial t}\frac{\partial\xi}{\partial r} 
  - \frac{1}{2}\frac{\partial\xi}{\partial
  t}\frac{\partial\xi}{\partial r} = 0 
  \ , \label{eq:metric1} \\ \bs
  \fl 2\rme^\xi - \rme^{2\upsilon}\left[
    \frac{\partial^2\xi}{\partial r^2}
    + 2\frac{\partial^2\chi}{\partial r^2}
    - 2\left( \frac{\partial\chi}{\partial r} \right)^2
    + \frac{\partial\chi}{\partial r}\frac{\partial\xi}{\partial r}
    + \frac{\partial\upsilon}{\partial r}%
      \frac{\partial\xi}{\partial r}
      + 2\frac{\partial\upsilon}{\partial r}%
      \frac{\partial\chi}{\partial r} \right] \nonumber \\ \ms
    \fl\qquad - \rme^{2\chi}\left[
    \frac{\partial^2\xi}{\partial t^2}
    - 2\frac{\partial^2\upsilon}{\partial t^2}
    + 2\left( \frac{\partial\upsilon}{\partial t} \right)^2
    - \left( \frac{\partial\xi}{\partial t}\right)^2
    + \frac{\partial\upsilon}{\partial t}\frac{\partial\xi}{\partial t}
    + \frac{\partial\chi}{\partial t}%
      \frac{\partial\xi}{\partial t}
      - 2\frac{\partial\chi}{\partial t}%
      \frac{\partial\upsilon}{\partial t} \right] = 0 \ ,
      \label{eq:metric2} 
\end{eqnarray}
respectively. Here $\xi \deq \ln K$, $\chi \deq  \ln X$ and $\upsilon
\deq \ln y$. 
  
These are essentially the equations one would obtain from a metric
ansatz, \cite{Exact}, since the functions $X$, $y$ and $K$ correspond
to the metric coefficients (see section \ref{sec:Isometry}).

Non-comoving coordinates has been used by some authors
(see \cite{McVittie,Knutsen1,Knutsen2,Knutsen3,Knutsen4,Knutsen5,%
  Bonnor-Knutsen} and references therein)
to analyze spherically symmetric perfect
fluids (see also \cite{Exact}). If the line-element is to remain
diagonal, we must have $xX - yY = 0$ (see section
\ref{sec:Isometry}). In general, this is a too weak constraint to yield
any significant simplifications of the equations, and further
assumptions has to be made.

We now continue presenting some of the special cases appearing in this
class of space-times.

The \emph{shear- and expansion free fluid} space-times have
$\al = \be = 0$, and they are therefore a bit pathological because of
their time-dependence (although the static
subcase is of great physical interest). In fact, the equations
(\ref{eq:ricci1'})--(\ref{eq:ricci5'}) and
(\ref{eq:bianchi1'})--(\ref{eq:bianchi2'}) reduce to
\begin{eqnarray}
  \bfe_1(a) = \tfrac32K - \tfrac12\mu - \tfrac32\kappa^2 - a\kappa +
  a^2 \ , \label{eq:stat21} \\
  \bfe_1(\kappa) = -\tfrac12K + \tfrac12\mu + \tfrac32\kappa^2 \ ,
  \label{eq:stat22} \\
  \bfe_1(K) = 2\kappa K \ , \label{eq:stat23}
\end{eqnarray}
(which are just the static set of equations when $x$ is
zero), with $p = -K + 2a\kappa + \kappa^2$, and
\begin{eqnarray*}
  \bfe_0(p) = 2\kappa\bfe_0(a) \ , \\
  \bfe_0(\kappa) = \bfe_0(K) = \bfe_0(\mu) = 0 \ , 
\end{eqnarray*}
where the last equality follows from (\ref{eq:contrBian1}). If we
align our spatial coordinate so that $x = 0$, we see that all the
time-dependence appears in $p$, $a$, and the gradients of the
coordinates. This is a manifestation of the
fact that this type of solution may be generated from a static
solution \cite{Exact}.

First, suppose that $K' = 0$. Then it is straightforward to show that
either $K = 0$ or $\mu + p = 0$.  
Therefore, if $K > 0$ (the case $K < 0$ can be treated in an analogous
  manner), we choose $K \deq 1/r^2$ $\Rightarrow$ $x = 0$, $y = -r\kappa$,
and we find that the commutator (\ref{eq:commutator2}) is identically
satisfied.
Assuming that $\mu$ is a given function of $r$, we may then integrate
equation (\ref{eq:stat22}) for $\kappa$. Alternatively we may define
$\mu$ through equation (\ref{eq:stat22}).

If $\bfe_0(a) = X\dot{a} \neq 0$, we can choose $a \deq t$, and $Y$ is
given by equation (\ref{eq:stat21}). Equation (\ref{eq:commutator1})
can then be integrated for $X$, since all other functions are known.

If $\bfe_0(a) = 0$ ($\Leftrightarrow$ $\bfe_0(p) = 0$), all the
rotation coefficients and curvature components depend solely on
$r$. Thus all the time-dependence enters through the coordinate
gradients. This is another example of when the coordinate gradients
contain essential information about space-time.
If $\dot{X} \neq 0$, we take $t = X$, and $Y$ can be obtained from
(\ref{eq:commutator1}) since $a$ is in principle known from equation
(\ref{eq:stat21}). If $\dot{X} = 0$ but $\dot{Y} \neq 0$, the
analogous result holds.

\emph{Shear free fluids} are characterized by $\alpha = \beta$.
If we transform our coordinates such that $Y = 0$, equation
(\ref{eq:ricci3'}) gives $\beta = \beta(t)$. If we make one further
coordinate transformation, to obtain $x = 0$, we can use equations
(\ref{eq:metric1}) and (\ref{eq:metric2}) as our field equations. The
shear-free condition can then be stated as $\upsilon = \xi/2 + g(r)$,
where $g$ is an integration function. Further, equation
(\ref{eq:metric1}) can be integrated so that $\chi = -\ln\dot\xi +
f(t)$, where $f$ in an integration function. Thus, equation
(\ref{eq:metric2}) becomes an equation solely for $\xi$, and all the
time-derivatives in the last bracket drop out. We define our
spatial coordinate $r$ by choosing a special form of the function $g$.

\subsection{Acceleration free fluids (including dust)}

For constant pressure the equations may be completely integrated, as
is indicated in \cite{Exact}.
If one chooses $Y=x=0$ it follows from (\ref{eq:contrBian2}) that
the pressure
is a function of $t$ only. From (\ref{eq:commutator2}) one gets that
also
$X$ only depends on $t$. We then still have the freedom to make $X=1$
by performing a coordinate transformation $t'=t'(t)$. This is in
accordance with the possibility to choose a synchronous
and comoving frame for non-accelerated matter \cite{Landau}. Note
that this choice could contradict an arbitrary choice of coordinates
from $R^{p+1}$. From equations (\ref{eq:bianchi1'}) and
(\ref{eq:ricci4'})
it then easily follows that
\[
  K=\kappa^2f^2(r),
\]
where $f(r)$ is an arbitrary function of $r$. The subsequent solving
of the equations is greatly simplified if we introduce the function
$S$ by
\[
  S \deq -\frac{1}{\kappa f}. 
\]
As will be seen in the last section, $S^2$ will then appear in the
metric in front of $\der\Omega^2$.
The remaining quantities may then be expressed in terms of $S$ as
\[
  y=\frac{1}{fS'} \ , \ \   
  \alpha=\frac{\dot S'}{S'} \ , \ \ \beta=\frac{\dot S}{S} \ ,
  \]
\[
  \mu = 2\frac{f'}{f^3SS'} + \frac{1}{S^2} - \frac{1}{f^2S^2} + 
        2\frac{\dot S \dot S'}{SS'} + \frac{{\dot S}^2}{S^2} \ ,
\]
and $S$ is given by the following equation
\[
  {\ddot S}S + \frac{{\dot S}^2}{2} + \frac{1}{2} - \frac{1}{2f^2} +
  \frac{pS^2}{2} = 0 \ ,
\]
(see also \cite{Exact}). 
The pressure $p = p(t)$ is free to specify.
If $p$ is a constant one may solve for $t$ as a function of $S$
and $r$. The
solution is given by
\[
  t(S,r) = \pm \int^S \sqrt{\frac{\hat S}{\hat{S}(f^{-2} - 1) + g - 
  \tfrac13p\hat{S}^3}}\ \der\hat{S} + h \ ,
\]
where $g$ and $h$ are arbitrary functions of $r$. If $p=0$ the
integral may in
fact be expressed in terms of elementary functions. Even if one cannot
solve for $S = S(t)$ in general, one can get an explicit line-element by
changing coordinate from $t$ to $S$. Off-diagonal terms will then of
course be introduced in the line-element.

\subsection{Fluids with all kinematical quantities non-zero}

Kitamura \cite{Kitamura} found a class of solutions with non-zero
shear, expansion, and acceleration, generalizing the results of
Gutman and Bespal'ko \cite{Gutman} (a solution rediscovered by Wesson
\cite{Wesson}), Lake \cite{Lake}, Van den Bergh and Wils
\cite{VandenBergh-Wils}, Collins and Lang \cite{Collins-Lang},
Coley and Tupper \cite{Coley-Tupper}, and Sussman \cite{Sussman}.
The solution of Kitamura can be obtained using the equations presented
here with the assumptions $x = Y = 0$, $X = -y$, $a = \kappa$, and
$\alpha = \alpha(t)$.

Using non-comoving coordinates, Knutsen \cite{Knutsen5} found a
spherically symmetric perfect fluid solution with shear, expansion and
acceleration. This solution is from a physical point of view quite
well behaved, but as shown in \cite{Knutsen5} it can not be matched to
a vacuum exterior. Also, it turns out that in general it is not
possible to use comoving coordinates to express the solution in terms
of ``well known analytic functions'' \cite{Knutsen5}.

Another solution with all kinematic quantities non-zero,
which is new to the authors knowledge, may be found with the
ansatz 
\[
  \begin{array}{ll}
    Y = x = 0 \ , & K = f(t)/r^2 \ , \\
    \alpha = 0 \ , & a = \kappa  \ .
  \end{array}
\]
The form of $K$ defines the coordinate $r$,
while retaining some freedom in the choice of time-coordinate.

From equation (\ref{eq:ricci4'}) we see that $\kappa = \kappa(r)$, and
from equations (\ref{eq:bianchi2'}) and (\ref{eq:commutator1}) we find
\[
  y = -\kappa r \ , \quad X = \frac{g(t)}{r} \ .
\]
where $g$ is an integration function. 
Further, equation (\ref{eq:ricci3'}) can be integrated so that 
\[
  \beta = \frac{F(t)}{r} \ ,
\]
and inserting this into the Bianchi equation (\ref{eq:bianchi1'}) we 
have
\begin{equation}\label{eq:feq}
  g\dot{f} + 2Ff = 0 \ ,
\end{equation}
which we use to define $g$. Expression for $p$ and $\mu$ can be
obtained from equation (\ref{eq:ricci1'}) and (\ref{eq:ricci2'})
respectively: 
\[
p = \frac{3f}{r^2} - 2\mu - 3\kappa^2 + \frac{3F^2}{r^2} \ , \quad
\mu = \frac{2f}{r^2} - 3\kappa^2 + \frac{3F^2}{r^2} +
  \frac{g\dot{F}}{r^2} \ .
\]
Inserting this expression for $\mu$ in equation (\ref{eq:ricci5'}), we
can integrate and find
\[
\kappa^2 = \frac{\Omega}{2}\frac{1}{r^2} - B \ , \quad
F^2 = Cf^2 - f + \frac{\Omega}{2} \ ,
\]
where $\Omega$, $B$ and $C$ are constants.
With this, $\mu = p + 6B$.
$f$ may be chosen as the time coordinate $t$.

We also see that $\mu = Cf^2/r^2 + 3B$, so that
$\mu$ is monotonically decreasing with increasing $r$, as is $p$.
The fluid as such is thus reasonably well behaved, although it
harbours a singularity at $r = 0$, but unfortunately it
can not be matched to the exterior Schwarzschild solution. The reason
for this is the following: Among the junction condition
(using the Darmois--Israel matching procedure \cite{Darmois,Israel})
on the matching surface are $a = a_{\rm ext}$ and $\kappa =
\kappa_{\rm ext}$, where we have used the subscript ``ext'' to denote
the corresponding quantities for the exterior solution. For the
particular fluid solution presented here (and also for the solution of
Kitamura), we have $a = \kappa$. Since this should hold on the
junction surface, we have $a_{\rm ext} = \kappa_{\rm ext}$ on that
surface. If we denote
Scwarzschild's radial coordinate by $R$, the latter condition gives a
constant matching radius $R = 3m$ (where $m$ is the mass parameter of
the Schwarzschild solution). On the other hand, from the remaining
matching conditions on finds that the matching radius is a
time-dependent quantity, thus giving a contradiction. This problem
also arises in Knutsen's solution. 

It is straightforward to show that Kitamura's solution 
and the solution presented in this section are not generically 
equivalent. Equivalence can be obtained for certain values of the
integration constants, but with arbitrary values of the integration
constants in the two solutions there is no equivalence. Also, our
solution is not equivalent to Knutsen's.

\subsection{Space-times with a homothetic motion}

If space-time is self-similar, i.e., has a homothetic Killing vector
$\xi$ the field equations
(for the LRS case) are reduced to ordinary differential equations (see
\cite{Carr-Coley2} for a review).
This is not the only reason for studying these space-times. They often
contain interesting physical phenomena like, e.g., shock waves, black
hole formation, evolution of voids, possible violation of cosmic
censorship, etc.\ \cite{Ori-Piran,Bogo2,Carr-Yahil,Carr-Hawking,%
  Bicknell-Henriksen,Evans-Coleman,Maison,Koike-Hara-Adachi}.
Space-time has a homothetic Killing vector if
\[
  \xi_{\mu;\nu}+\xi_{\nu;\mu}=2cg_{\mu\nu} ,
\]
where the constant $c$ may be chosen to $c=1$. If the coordinate $u$
is adapted to the homothetic vector, so that $\xi = \partial_u$, it is
easy to show that all components of the metric will depend on $u$ as
$g_{\mu\nu}=\rme^{2u}g^0_{\mu\nu}$, where $g^0_{\mu\nu}$ is independent of
$u$. It then follows that the components of the frame vectors $\bfe_i$
depend on $u$ as $\rme^{-u}$. The Ricci rotation coefficients
$\gamma_{ijk}$ also depend on $u$ as $\rme^{-u}$
and the Riemann tensor $R_{ijkl}$ as $\rme^{-2u}$
(see also \cite{RosJan}). Inserting
this into the IC (\ref{eq:commutator1})--(\ref{eq:bianchi2}) gives a
system of ordinary differential equations.
Note that we no longer can put $x=Y=0$ in general, since this could
contradict the choice $\xi = \partial_u$. However, it is still
possible to perform a coordinate transformation $u'=u+f(v)$, where
$v$ is the other essential coordinate, giving $\xi=\partial_{u'}$.
Hence the functional dependence on $u'$ will be the same as on $u$.
With such a transformation it is possible to remove $Y$ in the
time-like case and $x$ in the space-like.

Concerning the equation of state we have the following well known
result: If we assume that the equation of state is barotropic, i.e.,
$p = p(\mu)$ (which is reasonable in the case of a perfect fluid), the
homothetic motion gives $p = (\gamma - 1)\mu$, where $\gamma$ is a
constant. 

\subsubsection*{Temporal self-similarity}

Here $t=u$. In  general the system still gets quite complicated:
\begin{eqnarray*}
  \beta' = \frac{\kappa}{y}(\beta - \alpha) \ , \\ \bs
  \kappa' = \frac{\kappa^2}{y} + \frac{a\beta}{x} \ , \\ \bs
  \left(\frac{x}{y}\right)' = \frac{xa}{y^2} - \frac{x\kappa}{y^2}
  + \frac{\alpha}{y} - \frac{\beta}{y} \ , \\ \bs
  x\alpha' + ya' = \mu - 3\frac{y}{x}a\beta + a^2 - \alpha^2 -a\kappa
  - 2\alpha\beta \ ,
\end{eqnarray*}
supplemented by the defining equations
\begin{eqnarray*}
  X = \beta + \frac{x\kappa}{y} \ , \\
  K = \mu + \kappa^2 -\beta(2\alpha + \beta) -2\frac{y}{x}a\beta \ ,
  \\ \bs
  \mu = \frac{2}{\gamma}\left(a\kappa + \alpha\beta +
    \frac{y}{x}a\beta + \frac{x}{y}\kappa\alpha\right) \ , \\ \bs
  a\frac{\gamma}{\gamma - 1} + \alpha\frac{y}{x}\gamma = 2\kappa -
  2\frac{y}{x}(\gamma - 1)\beta \ .
\end{eqnarray*}

When rewritten in terms of dimensionless variables (chosen such that
the phase space is compact), this system (and the corresponding
spatial one) has been studied as a dynamical system in
\cite{Goliath-Nilsson-Uggla1,Goliath-Nilsson-Uggla2} (see also
\cite{Carr-Coley1}).

We here look at some simple special cases.
The assumption $\alpha=\beta$ (no shear) gives $\beta_0 \deq
\beta\rme^t =$ constant, $a = 0$, $K = \kappa^2$ and $\mu =
3\beta_0^2\rme^{-2t}$. If one chooses the coordinate $r$ as
$r = x/y$ one gets 
\[
  \kappa = -\frac{y}{r} = -\frac{x}{r^2} = \frac{(3\gamma -
  2)\beta_0}{2r}\,\rme^{-t}  \ , \ \
  X = \frac{3\beta_0\gamma}{2}\,\rme^{-t} .
\]
The pressure is given by $p = (\gamma - 1)\mu$ and all values in the
interval $1\leq \gamma \leq 2$ are allowed.

The simplification $x=0$ leads to $K=0$ (otherwise one gets $\mu=0$).
If one chooses the space-like coordinate $r$ as $r=\beta\rme^t$
the solution is given by
\[
  X = \frac{2\gamma}{2 - \gamma}r\rme^{-t} \ , \ \
  y = \frac{2 - 3\gamma}{2 - \gamma}r\kappa_0\rme^{-t} \ , \ \   
  \alpha = \frac{2\gamma}{2 - \gamma}r\rme^{-t} \ , \ \
  \beta=r\rme^{-t} \ ,
\]
\[
  a = \frac{2 - 3\gamma}{2 - \gamma}\kappa_0\rme^{-t} \ , \ \
  \kappa = \kappa_0\rme^{-t} \ , \ \
  \kappa_0^2 = r^2 - Dr^{\frac{\gamma}{\gamma - 1}} \ , 
\]
and
\[
  p = (\gamma -1)\mu =
  \frac{7\gamma - 6}{2 - \gamma}Dr^{\frac{\gamma}{\gamma-1}}%
  \rme^{-2t} \ , \ \
\]
where $D$ and $\gamma$ are constants. We see that the allowed
interval for $\gamma$ is given by $1<\gamma<2$.
The 1-forms are given by (see also section \ref{sec:Isometry} on how
to obtain $\w^2$ and $\w^3$)
\begin{eqnarray*}
  \w^0 = \frac{2 - \gamma}{2\gamma}\frac{\rme^t}{r}\,\der{t} \ , \quad  
  \w^1 = \frac{2 - \gamma}{2 - 3\gamma}\frac{\rme^{t}}{\kappa_0 r} 
  \, \der{r}  \ , \\ \bs
  \w^2 = r^{\frac{\gamma - 2}{2 - 3\gamma}}\rme^{\frac{2 -
  \gamma}{2\gamma}t} \, \der{v} \ , \quad \w^2 = r^{\frac{\gamma -
  2}{2 - 3\gamma}}\rme^{\frac{2 - \gamma}{2\gamma}t} \, \der{w} \ .
\end{eqnarray*}
Note that in these coordinates all the components $g_{\mu\nu}$ of the
metric are not proportional to $\rme^{2t}$. This proportionality may
however be achieved by a coordinate transformation of the
non-essential coordinates $v$ and $w$.

\subsubsection*{Spatial self-similarity}

Here $r=u$. These space-times can be analyzed analogously to the temporal case.
As an example of an exact solution, we look at the case when
$Y=0$. This implies
$\kappa =  a = -y$, $\alpha = 0$, and $\mu = p$. We find that $\kappa$
is a constant, and $K = \mu + \kappa^2 - \beta^2$. If $\mu\rme^{2r}$ is
chosen as the time-like coordinate one obtains
\[
  \beta^2 = \mu + \kappa^2 - C\mu^{1/2}\,\rme^{-r}  \ ,
\]
where $C$ is an integration constant.

\section{The isometry algebra}\label{sec:Isometry}

The isometry algebra may be obtained by projecting the 1-forms onto
the orbits by putting $x^{\alpha} =$ constant and all generators of
the Lorentz group, $\tau^i\!_j$,
except $\tau^2\!_3$ to zero.
Defining ${\w}^4 \deq {\w}^2\!_3 =  {\tau}^2\!_3 +
{\gamma}^2\!_{3i}\omega^i$, the isometry algebra on the
orbits is given by (by $|$ we denote projection onto the orbits) 
\begin{eqnarray}
  k\der{\w} = 0 \ , \label{eq:iso1}\\
  \der{\w}^2| = {\w}^3|\wedge{\w}^4| + F{\w}\wedge{\w}^2|
  \ , \label{eq:iso2}\\
  \der{\w}^3| = {\w}^4|\wedge{\w}^2| + F{\w}\wedge{\w}^3|
  \ , \label{eq:iso3}\\
  \der{\w}^4| = K{\w}^2|\wedge{\w}^3| \ , 
  \label{eq:iso4}
\end{eqnarray}
where ${\w}$, $F$, and $k$ are defined in Table 1.

\begin{table}
  \caption{Definitions of the quantities used in
    section \protect{\ref{sec:Isometry}}.}
  \begin{indented}
  \item[]\begin{tabular}{lllll}
  \br
   Case    & ${\w}$      & $F$       & $k$   & $\xi$ \\
  \mr
   Static  & ${\w}^{0}|$ & $\beta$   & $1$   & $t$, time-like  \\
   Spatially homogeneous      & ${\w}^1|$   & $-\kappa$ & $1$   & $r$,
  space-like \\
   Inhomogeneous and time dependent & ---         & ${0}$     & ${0}$
  & ---        \\ 
  \br
\end{tabular}
\end{indented}
\end{table}

\subsection{Spherical and hyperbolic symmetric space-times
  $(K \neq 0)$}

When $K \neq {0}$ the integrability conditions give $F = {0}$ (see
sections \ref{sec:static} and \ref{sec:homo}). Depending on the value
of ${\sgn}(K)$ the algebra (\ref{eq:iso2})--(\ref{eq:iso4})
has two different solutions:
\begin{enumerate}
  \item[(a)] ${\sgn}(K) = 1$. 
    \begin{eqnarray*}
      {\w}^2| = K^{-1/2}\left(-\sin\zeta\, \der\theta +
      \sin\theta\cos\zeta\, \der\phi\right) \ , \\
      {\w}^3| = K^{-1/2}\left(\cos\zeta\, \der\theta +
      \sin\theta\sin\zeta\, \der\phi\right) \ , \\
      {\w}^4| = \cos\theta\, \der\phi + \der\zeta \ .
    \end{eqnarray*}
  \item[(b)] ${\sgn}(K) = -1$. 
    \begin{eqnarray*}
      {\w}^2| = |K|^{-1/2}\left(-\sin\zeta\, \der\theta +
      \sinh\theta\cos\zeta\, \der\phi\right) \ , \\
      {\w}^3| = |K|^{-1/2}\left(\cos\zeta\, \der\theta +
      \sinh\theta\sin\zeta\, \der\phi\right) \ , \\
      {\w}^4| = \cosh\theta\, \der\phi + \der\zeta \ .
    \end{eqnarray*}
\end{enumerate}

For the static and SH cases, equation (\ref{eq:iso1}) has the solution
${\w} = \der\xi$.

The line elements are then given by\\[2mm]
\textit{Static case}
\[
  {\met} = \left\{
    \begin{array}{ll}
      A^2\,{\der}t^2 - y^{-2}\,{\der}r^2 - K^{-1}\left(\der\theta^2
      + \sin^2\theta\der\phi^2\right) \ , & K > 0 \\ \bs
      A^2\,{\der}t^2 - y^{-2}\,{\der}r^2 - |K|^{-1}\left(\der\theta^2
        + \sinh^2\theta\der\phi^2\right) \ , & K < 0 \\
    \end{array} \right. \ ,
\]
where
\[
  A(r) = A_0\rme^{-\int^r(a/y)\,{\der}\hat{r}} \ ,
\]
with $A_0$ a constant;\\[2mm]
\textit{SH case}
\[
  {\met} = \left\{
    \begin{array}{ll}
      X^{-2}\,{\der}t^2 - B^2\,{\der}r^2 - K^{-1}\left(\der\theta^2
      + \sin^2\theta\der\phi^2\right) \ , & K > 0 \\ \bs
      X^{-2}\,{\der}t^2 - B^2\,{\der}r^2 - |K|^{-1}\left(\der\theta^2
        + \sinh^2\theta\der\phi^2\right) \ , & K < 0 \\
    \end{array} \right. \ ,
\]
where
\[
  B(t) = B_0\rme^{\int^t(\alpha/X)\,{\der}\hat{t}} \ ,
\]
with $B_0$ a constant;\\[2mm]
\textit{Generic case}
\[
  {\met} = \left\{
    \begin{array}{rl}
      \begin{array}{ll}
        \Delta^{-2}\left[(y^2 - x^2){\der}t^2 + 2(xX -
        yY){\der}t{\der}r - (X^2 - Y^2){\der}r^2
      \right] & \\
      \quad- K^{-1}\left(\der\theta^2
        + \sin^2\theta\der\phi^2\right) \ , 
      \end{array} & K > 0\\
      \begin{array}{ll}
        \Delta^{-2}\left[(y^2 - x^2){\der}t^2 + 2(xX -
        yY){\der}t{\der}r - (X^2 - Y^2){\der}r^2
      \right] & \\
      \quad- |K|^{-1}\left(\der\theta^2
        + \sinh^2\theta\der\phi^2\right) \ , 
      \end{array} & K < 0
    \end{array} \right.
\]
where $\Delta \deq yX - xY$.

\subsection{Plane symmetric space-times $(K = 0)$}

If ${\w} = \der\xi$, then the algebra
(\ref{eq:iso2})-(\ref{eq:iso4}) has the solution
\begin{eqnarray*}
  {\w}^2| = \mathrm{e}^{F\xi}\left( \cos\zeta\,\der v -
  \sin\zeta\,\der w \right) \ , \\
  {\w}^3| = \mathrm{e}^{F\xi}\left( \sin\zeta\,\der v +
  \cos\zeta\,\der w \right)\ , \\
  {\w}^4| = \der\zeta \ ,
\end{eqnarray*}
where $v$ and $w$ are space-like. Putting $F = {0}$ we obtain the
solution to the generic algebra.

The metrics become\\[2mm]
\textit{Static case}
\[
  {\met} = A^2{\der}t^2 - y^{-2}({\der}r - xA\,{\der}t)^2 -
  \rme^{2B}\left({\der}v^2 + {\der}w^2\right) \ ,
\]
where
\begin{eqnarray*}
  A(r) \deq C_s\rme^{-\int^r(a/y)\,{\der}\hat{r}} \ , \\
  B(t,r) \deq -\int^r\frac{\kappa}{y}\,{\der}\hat{r} + D_st \ ,
\end{eqnarray*}
with $C_s$ and $D_s$ constants;\\[2mm]
\textit{SH case}
\[
  {\met} = X^{-2}({\der}t - YG\,{\der}r)^2 - G^2{\der}r^2 -
  \rme^{-2H}\left({\der}v^2 + {\der}w^2\right) \ ,
\]
where
\begin{eqnarray*}
  G(t) \deq C_h\rme^{\int^t(\alpha/X)\,{\der}\hat{t}} \ , \\
  H(t,r) \deq -\int^t\frac{\beta}{X}\,{\der}\hat{t} + D_hr \ , \\
\end{eqnarray*}
with $C_h$ and $D_h$ constants;\\[2mm]
\textit{Generic case}
\begin{eqnarray*}
  {\met} &=& \Delta^{-2}\left[(y^2 - x^2){\der}t^2 + 2(xX -
        yY){\der}t{\der}r - (X^2 - Y^2){\der}r^2
      \right] \nonumber \\
      && \quad - \rme^{2f}\left({\der}v^2 + {\der}w^2\right) \ ,
\end{eqnarray*}
where $f = f(t,r)$ satisfies
\[
{\bfe}_0(f) = \beta \ , \quad {\bfe}_1(f)= -\kappa \ ,
\]
which is a set of integrable equations due to the integrability
conditions. 

\ack

We would like to thank H.\ Stephani for pointing out some invaluable
references. We also thank K.\ Markstr\"om and F.\ St{\aa}hl at the
Department of Mathematics, Ume{\aa} University, for helpful
discussions. This work was supported by the Swedish Natural Science
Research Council. 


\end{document}